\documentclass[matbio]{svjour}
\usepackage{t1enc}
\usepackage{amssymb,amsmath,graphicx}
\usepackage[latin1]{inputenc}
\usepackage[english]{babel}
\newcommand{\fra}[2]{\textstyle{\frac{#1}{#2}}}
\newcommand{\beqn}{\begin{eqnarray}\begin{aligned}}
\newcommand{\eqn}{\end{aligned}\end{eqnarray}}

\begin{document}

\title{Entanglement Invariants and Phylogenetic Branching}
\author{J G Sumner$^*$\thanks{$^*$ Australian Postgraduate Award} \and P D Jarvis$^\dagger$\thanks{$^\dagger$ Alexander von Humboldt Fellow}}
\institute{School of Mathematics and Physics, University of Tasmania GPO Box 252-21, Hobart Tas 7001, Australia. \email{jsumner@utas.edu.au}}
\maketitle
\abstract{It is possible to consider stochastic models of sequence evolution in phylogenetics in the context of a dynamical tensor description inspired from physics. Approaching the problem in this framework allows for the well developed methods of mathematical physics to be exploited in the biological arena. We present the tensor description of the homogeneous continuous time Markov chain model of phylogenetics with branching events generated by dynamical operations. Standard results from phylogenetics are shown to be derivable from the tensor framework. We summarize a powerful approach to entanglement measures in quantum physics and present its relevance to phylogenetic analysis. Entanglement measures are found to give distance measures that are equivalent to, and expand upon, those already known in phylogenetics. In particular we make the connection between the group invariant functions of phylogenetic data and phylogenetic distance functions. We introduce a new distance measure valid for three taxa based on the group invariant function known in physics as the "tangle". All work is presented for the homogeneous continuous time Markov chain model with arbitrary rate matrices.}

\section{Introduction}

Stochastic methods which model character distributions in aligned gene sequences are part of the standard armoury of phylogenetic analysis \cite{stee,fels,fels2,rodr,nei}. The evolutionary relationships are usually represented as a bifurcating tree directed in time. It is remarkable that there is a strong conceptual and mathematical analogy between the construction of phylogenetic trees using stochastic methods, and the process of scattering in particle physics \cite{jarv}. It is the purpose of the present work to show that there is much potential in taking an algebraic, group theoretical approach to the problem where the inherent symmetries of the system can be fully appreciated and utilized.\\\indent
Entanglement is of considerable interest in physics and there has been much effort to elucidate the nature of this most curious of physical phenomena \cite{wern,lind,bern,dur,guhn}. Entanglement has its origin in the manner in which the state probabilities of a quantum mechanical system must be constructed from the individual state probabilities of its various subsystems. Whenever there are global conserved quantities, such as spin, it is the case that there exist entangled states where the choice of measurement of one subsystem can affect the measurement outcome of another subsystem no matter how spatially separated the two subsystems are. This curious physical property is represented mathematically by nonseparable tensor states. Remarkably, if the pattern frequences of phylogenetic analysis are interpreted in a tensor framework it is possible to show that the branching process itself introduces entanglement into the state. This is a mathematical curiosity that can be studied using methods from quantum physics. This is a novel way of approaching phylogenetic analysis which has not been explored before. \\\indent
In section \ref{tensor} we begin by considering the stochastic model of sequence evolution in phylogenetics using continuous time Markov chains (CTMCs)\footnote{The reader should note that the model considered in this paper is the general Markov model on a tree together with the additional assumption that the transition matrices are non-singular and arise as an analytical continuation from the identity matrix. For a recent review of the hierarchy of phylogenetic models see \cite{erik}.}. We go on to present this model in a dynamical tensor description where the probability distribution is given by the components of a tensor in a preferred basis and the time evolution is generated by linear operators acting on the space. The phylogenetic branching process is then developed formally in section \ref{branching} by introducing a linear operator which introduces an extra product in the tensor space. This operator is shown to be unique given that the probability distribution must be conditionally independent from branch to branch. We also show that the branching process introduces entanglement into the state space. The stationary states of the system and the pulley principle, which describes the unrootedness of phylogenetic trees, are presented in the tensor framework in sections \ref{stationarysection} and \ref{pulley} respectively. Section \ref{orbitssection} is a short review of current methods of analysing entanglement in terms of group orbits and invariant functions.  In sections \ref{2characters} and \ref{canonical} the work specializes to the cases of two phylogenetic characters and small numbers of taxa making up the analysis. In section \ref{concurrencesection} the group orbits and invariant functions for the case of two taxa are presented and explicitly solved to show that the invariant function is the well known $\log\det$ distance. In section \ref{tangle} we go on to study the case of three taxa where the invariant function known in physics as the tangle is shown to give a new distance measure for phylogenetics. This previously unstudied distance measure is found to be  useful analytical tool in the reconstruction of phylogenetic trees, (section \ref{distances}).

\section{Tensor methods in phylogenetic branching}\label{tensor}
We consider a system consisting of $N$ sites each of which takes on one of $K$ distinct characters. Associated with such a system we have the set of frequencies
\beqn
\widehat{p}_i:&=\frac{\text{total number of occurrences of character }i}{N},\\i&=0,1,...,K-1.\nonumber
\eqn
We model these frequencies by defining a set of probabilities which are the theoretical limit
\beqn
p_i=\lim_{N\rightarrow\infty}\widehat{p}_i.\nonumber
\eqn
 Introducing the $K$ dimensional vector space $V$ with preferred basis \{$e_i$\}, we can associate the set of probabilities with a unique vector
\beqn
p=p_0e_0+p_1e_1+...+p_{K-1}e_{K-1}.\nonumber
\eqn  
The probabilities are assumed to evolve in time as a homogeneous CTMC \cite{hagg,rind}. This amounts to assuming that the character
state at any time $t$, conditional on the character
 state any time $t' < t$, is independent of the
 character state at any earlier time $t'' < t'$
\cite{fels,fels2,rodr}. The defining relation for the time evolution is  
\beqn
\frac{d}{dt}p(t)=R\cdot p(t),\label{rate}
\eqn
where $R$ is a linear operator. To preserve reality of the probabilities and the property $\sum_i p_i(t)=1,\forall t$ it follows that in the preferred basis $R$ is a real valued zero column sum matrix. A formal solution of (\ref{rate}) for time independent $R$ is found by exponentiating
\beqn\label{exp}
p(t)=&e^{Rt}p(0)\\
:=&M(t)p(0).
\eqn
We refer to $M(t)$ as a Markov operator. Taking its derivative
\beqn
\frac{d}{dt}M(t)=RM(t),\nonumber
\eqn
we observe that 
\beqn
\frac{d}{dt}M(t)|_{t=0}=R.\nonumber
\eqn
As is well known, in order to conserve positivity of the probabilities it must also be the case that
\beqn
R_{ij}&\geq0,\quad\forall i\neq j;\nonumber\\
R_{ii}&\leq0.
\eqn
\\\indent
In phylogenetics we consider the case where we have multiple, aligned, $N$ site, $K$ character systems labelled by $\{1,2,...,L\}$. We refer to the individual systems as taxa. What is now of interest is the set of frequencies 
\beqn
\widehat{p}_{i_1i_2...i_L}:=&\frac{\text{total number of occurrences of pattern }i_1i_2...i_L}{N},\nonumber\\
i_1,i_2,...,i_L=&0,...,K-1.\nonumber
\eqn
We model these frequencies by again defining a set of probabilities which are the theoretical limit 
\beqn
p_{i_1i_2...i_L}:=\lim_{N\rightarrow\infty}\widehat{p}_{i_1i_2...i_L}.\nonumber
\eqn
The system is assumed to have evolved in time as a homogeneous CTMC.\\\indent
Introducing the random variables $x_1,x_2,...,x_L$ each of which take on values in the  individual character spaces $\{i_1,i_2,...,i_L=0,...,K-1\}$ and $x=(x_1,x_2,...,x_L)$ which takes on values in the $K^L$ dimensional character space $\{i_1i_2...i_L\}$ we can write the transition probabilities of the Markov chain as
\beqn
\mathbb{P}&(x\!=i_1i_2...i_L,t|x\!=\!j_1j_2...j_L,0)\nonumber\\&=\mathbb{P}(x_1\!=\!i_1,t|x_1\!=\!j_1,0)\mathbb{P}(x_2\!=\!i_2,t|x_2\!=\!j_2,0)...\mathbb{P}(x_L\!=\!i_L,t|x_L\!=\!j_L,0)\nonumber\\
:\!&=M^1_{i_1j_1}(t)M^2_{i_2j_2}(t)...M^L_{i_Lj_L}(t),\nonumber\\
&=\sum_{k_1,k_2,...,k_L}M^1_{i_1k_1}(t)M^2_{i_2k_2}(t)...M^L_{i_Lk_L}(t)\delta^{k_1}_{j_1}\delta^{k_2}_{j_2}...\delta^{k_L}_{j_L}.
\eqn
From this we notice that it is possible to construct the state space setting of the tensor product $V\otimes V\otimes ...\otimes V=V^{\otimes L}$ where the probabilities are associated with the tensor
\beqn
P(t)=\sum_{i_1,...,i_L}p_{i_1i_2...i_L}(t)e_{i_1}\otimes e_{i_2}\otimes ...\otimes e_{i_L}.\nonumber
\eqn 
Time evolution of this system is generated by the transition probabilities of the Markov chain which in tensor notation can be represented as linear operators acting on the initial pattern distribution
\beqn
P(t)=M^1(t)\otimes M^2(t)\otimes ...\otimes M^L(t)P(0),\nonumber
\eqn
where distinct rate parameters have been allowed on each component of the tensor product space. (The reader should note that $M^l$ refers to the $lth$ component of the tensor product space and is \textit{not} meant to indicate the $lth$ power of the operator $M$.)\\\indent
If we have a phylogenetic tensor $P(t)$ which describes the pattern distribution of characters for $L$ taxa, it is possible to find the reduced tensor $\overline{P}(t)$ which gives the pattern distribution for a subset of $l$ taken from the original set of $L$ taxa. The correct operation is given by
\beqn\label{reduced}
p_{i_1i_2...i_L}(t)&\rightarrow \overline{p}_{i_1i_2...i_l}(t)=\sum_{i_s,i_t,...}p_{i_1i_2...i_L}(t),\\
\overline{P}(t)&=\sum_{\text{all i's}}\overline{p}_{i_1i_2...i_l}(t)e_{i_1}\otimes e_{i_2}\otimes ...\otimes e_{i_l},
\eqn
where $s,t,...,$ label the taxa which are not in the subset. Such a reduced tensor will be referred to as a \textit{marginal} distribution. (In this work we will follow the convention from here on that if a summation sign has no suffixes it is assumed that \textit{all} indices inside the expression are to be summed over.)

\section{Phylogenetic Branching}
\label{branching}

\begin{figure}[h]
\centering
\resizebox{0.3\hsize}{!}{\rotatebox{270}{\includegraphics{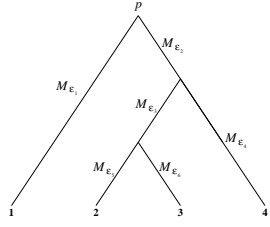}}}
\caption{The CTMC model of four taxa on the tree (1((23)4)).}
\label{fourtaxa}
\end{figure}

Having developed the general tensor description of the homogeneous CTMC model of sequence evolution in phylogenetics, in this section we will now introduce a formalism for describing the branching events. We do this by defining a formal operation on the tensor space. \\\indent
Consider the case where we have a single taxon branching into $L=2$ taxa. The corresponding mathematical operation is $V\rightarrow V\otimes V$. If the branching event occurs at $t=\tau$ we are required to determine the appropriate pattern probabilities $p_{i_1i_2}(\tau)$ given the probabilities $p_{i}(\tau)$. (In this paper $\tau$ is considered to be an additional model parameter alongside the parameters in the rate matrices). Intuitively a reasonable choice is the initial set $p_{ii}(\tau)=p_i(\tau)$ and $p_{ij}(\tau)=0,\forall i\neq j$. \\\indent
In order to formalize this we introduce the \textit{splitting} operator $\delta:V\rightarrow V\otimes V$. The most general action of $\delta$ on the basis elements of $V$ can be expressed as
\beqn\label{splitting}
\delta\cdot e_i=\sum_{j,k}\Gamma _i^{jk}e_j\otimes e_k,
\eqn
where $\Gamma _i^{jk}$ are an arbitrary set of coefficients. Standard models of phylogenetics assume conditional independence upon the distinct branches of the tree \cite{steel,fels,fels2,nei}. This assumption will presently be used to determine the exact form of the splitting operator. It is only necessary to consider initial probabilities of the form \beqn p^{(\gamma)}_{i}(\tau)&=\delta_i^{\gamma},\nonumber\\\gamma &=0,1,...,K-1\eqn so that the initial single taxon state is
\beqn
p^{(\gamma)}(\tau)&=\sum_{i}p^{(\gamma)}_i(\tau)e_i,\nonumber\\&=\sum_{i}\delta_i^\gamma e_i.\nonumber\\
\eqn
Directly subsequent to the branching event the 2 taxa state is given by
\beqn
P^{(\gamma)}(\tau)&=\delta\cdot p^{(\gamma)}(\tau),\nonumber\\
&=\sum_{i,j,k}\delta^\gamma_i\Gamma_i^{jk}e_j\otimes e_k.
\eqn
We implement the conditional independence upon the branches by setting
\beqn\label{condindep}
\mathbb{P}(x\!=\!i_1i_2,&t\!=\!t'|x_1\!=\!x_2=\gamma,t\!=\!\tau)\\
=\mathbb{P}&(x_1\!=\!i_1,t\!=\!t'|x_1\!=\!\gamma,t\!=\!\tau)\mathbb{P}(x_2\!=\!i_2,t\!=\!t'|x_2\!=\!\gamma,t\!=\!\tau).
\eqn
Using the tensor formalism the transitions probabilites can be expressed as
\beqn
&\mathbb{P}(x_1=i_1,t=t'|x_1=\gamma,t=\tau)=\sum_{k}M^1_{i_1k}(t'-\tau)\delta_k^{\gamma},\nonumber\\
&\mathbb{P}(x_2=i_2,t=t'|x_2=\gamma,t=\tau)=\sum_{l}M^2_{i_2l}(t'-\tau)\delta_l^{\gamma},\nonumber\\
&\mathbb{P}(x=i_1i_2,t=t'|x_1=x_2=\gamma,t=\tau)\nonumber\\
&\hspace{100pt}
=\sum_{k,l,m}M^1_{i_1k}(t'-\tau)M^2_{i_2l}(t'-\tau)\delta_m^\gamma\Gamma_{m}^{kl}.\nonumber
\eqn
Implementing (\ref{condindep}) leads to the requirement that 
\beqn\label{splitcond}
\Gamma_\gamma^{kl}=\delta_k^\gamma\delta_l^\gamma
\eqn
and the basis dependent definition of the splitting operator
\beqn\label{delta}
\delta\cdot e_i=e_i\otimes e_i.
\eqn
The action on the components of a vector is such that
\beqn\label{comp}
p_{ij}(\tau)&=p_i(\tau),\quad i=j;\\
&=0,\quad i\neq j;\nonumber
\eqn
which is consistent with our intuitive guess. As will become apparent, the operation defined in (\ref{delta}) takes disentangled states into entangled states.\\\indent 
The splitting operator is an important structural element of the tensor description and its symmetry properties \cite{bashjarv} are intimately related to the existence of discrete transform methods for particular classes of phylogenetic model. More general forms of the conditions (\ref{splitcond}) can be envisaged under weaker assumptions than considered here. Finally, in the particle scattering picture for phylogenetic branching \cite{jarv} the splitting operator is implemented as an interaction term. For present purposes the utility of $\delta$ is that it allows us to write down a formal expression for a system which undergoes a branching event.  Suppose a system described by the tensor $P\in V^{\otimes L}$ undergoes a branching event on its $r^{th}$ branch, the new system is described by
\beqn\label{coords}
P\rightarrow(1_1\otimes 1_2\otimes...\otimes 1_{r-1}\otimes\delta\otimes 1_{r+1}\otimes...\otimes 1_L)P\in V^{\otimes L+1},
\eqn
where $1_s$ is the identity operator on the $s$th component of the tensor product space.
We introduce the convention that the tensor space is labelled so that under the action (\ref{coords}) the probabilities are given by
\beqn
p_{i_1i_2...i_L}\rightarrow p_{i_1i_2...i_ri_{r+1}i_{r+2}...i_{L}i_{L+1}}=p_{i_1i_2...i_ri_{r+2}...i_Li_{L+1}}\delta_{i_ri_{r+1}}.\nonumber
\eqn
We introduce parameter sets labelled on the edges of the tree by \{$\epsilon_a,a=1,2,...\}$ and defined as $\epsilon_a=\{\alpha_a,\beta_a,...,.;t_a\}$ to distinguish entries in the rate matrices and branch lengths on different edges. Given the solution (\ref{exp}) it should be noted that there is an edge scaling symmetry \beqn
\{\alpha_a,\beta_a,...\}&\rightarrow\{\lambda\alpha_a,\lambda\beta_a,...\},\\t_a&\rightarrow \lambda^{-1}t_a\nonumber
\eqn 
which leaves the model invariant. This symmetry is well known in the literature \cite{fels,fels2} and indicates that it is not possible to distinguish between a fast rate of evolution and a long time period of evolution. When the rate parameters $\{\alpha_a,\beta_a,...\}$ are identical on all edges of the tree, a "molecular clock" is said to be in operation. Under the circumstances of a molecular clock it is possible in principle to determine the time period. \\\indent
The edges of a tree are labelled using an away-from-the-root and left-to-right ordering convention. As an example the expression which defines the most general homogeneous CTMC on the tree $(1((23)4))$ is given by
\beqn
P&_{\tau_2\tau_3}(t)\nonumber\\
&=(M_{\epsilon_1}\!\otimes\!M_{\epsilon_5}\!\otimes\!M_{\epsilon_6}\!\otimes\!M_{\epsilon_4})1\!\otimes\!\delta\!\otimes\!1(1\!\otimes\!M_{\epsilon_3}\!\otimes\!1)1\!\otimes\!\delta(1\!\otimes\!M_{\epsilon_2})\delta\cdot p\nonumber
\eqn
where $p$ is the initial single taxon distribution, $\tau_2$,$\tau_3$ define the branching times and $t_1=t$, $t_4=t-\tau_2$, $t_5=t_6=t-\tau_2-\tau_3$, (see figure \ref{fourtaxa}).\\\indent
 In this work we will derive results which are independent of the particular rate parameters which occur in the Markov operators of the model. 

\section{Stationary states}
\label{stationarysection}

A stationary state of a homogeneous CTMC is defined as the vector, $\pi$, which satisfies
\beqn\label{stationary}
R\cdot\pi=0.
\eqn
The stationary state can be generalized to the case of a tensor, $\Pi\in V^{\otimes L}$, associated with a set of pattern probabilities which satisfies 
\beqn
R_1\otimes R_2\otimes...\otimes R_L\cdot\Pi=0\nonumber
\eqn
and has solution
\beqn
\Pi=\pi_1\otimes\pi_2\otimes...\otimes\pi_L.\nonumber
\eqn
It should be noted that $\Pi$ is a completely separable state and that \textit{any} state tends to the stationary state as $t\rightarrow \infty$ \cite{hagg}.\\

\section{The pulley principle}\label{pulley}

\begin{figure}[h]
\centering
\resizebox{0.3\hsize}{!}{\rotatebox{0}{\includegraphics{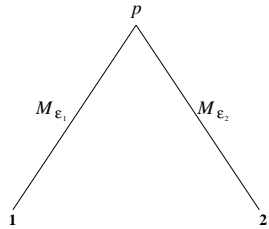}}}
\caption{The CTMC model of two taxa.}
\label{twotaxa}
\end{figure}

We consider the case of a single taxon which is in state $p\!\in\! V$ at time $t=0$. We implement a branching event at $t=0$ and let the system evolve under arbitrary Markov operators to produce the state at a later time $t$
\beqn\label{see}
P(t)=M_{\epsilon_1}\otimes M_{\epsilon_2}\delta\cdot p\quad\in\! V\otimes V,
\eqn
(see figure \ref{twotaxa}).

 The dual vector space $V^*$ has a basis which consists of the linear maps \{$f^i:V\rightarrow F,i=1,...,K$\} which satisfy $f^i(e_j)=\delta^i_j$. Using this dual basis we can define an isomorphism $\phi: V\otimes V\rightarrow L(V)$ as $\phi(e_i\otimes e_j)\rightarrow e_i f^j$ and rewrite (\ref{see}) as
\beqn
\phi(P(t))=M_{\epsilon_1}\phi(\delta\cdot p)M^{\intercal}_{\epsilon_2},\nonumber
\eqn
where ${\intercal}$ indicates matrix transpostion.\\\indent
The pulley principle is then a direct consequence of the existence of states $p$ such that $
M\phi(\delta\cdot p)=\phi(\delta\cdot p)M^\intercal$ 
for a given Markov operator $M$. A solution can be found to be 
\beqn
p&=\pi,\nonumber\\
\eqn
where $M\cdot\pi=\pi$ so that $\pi$ is a stationary state under $M$. Putting this together we can conclude that 
\beqn
P_1(t):=\left[M_{\epsilon_1}\otimes M_{\epsilon_2}\right]\delta\cdot\pi_1=&\left[1\otimes M_{\epsilon_2}M_{\epsilon_1}\right]\delta\cdot\pi_1,\nonumber
\eqn
or
\beqn
P_2(t):=\left[M_{\epsilon_1}\otimes M_{\epsilon_2}\right]\delta\cdot\pi_2=&\left[M_{\epsilon_1}M_{\epsilon_2}\otimes 1\right]\delta\cdot\pi_2,\nonumber
\eqn
for the special case where $M_{\epsilon_1}\cdot\pi_1=\pi_1$ or $M_{\epsilon_2}\cdot\pi_2=\pi_2$, respectively. This tells us that in the case where the initial distribution is a stationary state of one of the rate matrices on edges 1 and/or 2 the placement of the root of the tree is not strictly determined. This property has been observed previously to hold when the Markov chain is reversible, \cite{fels,fels2}. However in the tensor framework presented here we have refined the pulley principle by showing that one requires only that the initial distribution $\pi$ be a stationary distribution upon either branch of the tree for one to be able to "pull" that same branch through the initial distribution. This is a less stringent requirement than that of reversibility \cite{hagg}. \\\indent

\section{Group invariants and orbit classes}
\label{orbitssection}

In quantum physics there has been much interest in quantifying and/or classifying the phenomenon of entanglement between multiple non-local systems \cite{wern}. The correct description involves expressing the total state vector as belonging to the multi-linear space built from the tensor product of the individual state spaces. Entangled states exhibit non-local behaviour and correspond mathematically to the non-separable property of such state vectors.\\\indent
A systematic approach to the classification problem is to study the orbit classes of the tensor product space under a group action which is designed to preserve the essential non-local properties of entanglement. The orbit of an element $h$ belonging to the (multi)-linear space $H$ under the group action $G$ is defined as the set of elements $\{h'\in H:h'=gh\text{ for some }g\in G\}$.\\\indent
In quantum physics the appropriate group action is known to be the set of SLOCC operators, (Stochastic Local Operations with Classical Communication) \cite{lind,dur,guhn,nei2,miya}. Mathematically SLOCC operators correspond to the ability to transform the individual parts of the tensor product space $H\cong H_1\otimes H_2\otimes ...\otimes H_n$ with arbitrary invertible, linear operations. These operators are expressed by group elements of the form
\beqn
g=g_1\otimes g_2\otimes...\otimes g_n,\nonumber
\eqn
where $n$ is the number of individual spaces making up the tensor product, and $g_i\in GL(H_i)$.\\\indent
 The task is to identify the orbit classes of a given tensor product space under the general set of SLOCC operators. Powerful tools in this analysis are the methods of classical invariant theory. If $H$ is defined on a field $\mathbb{F}$, the set of invariant functions $I(G)$ is defined as
\beqn
I(G)=\{f:H\rightarrow\mathbb{F},\text{ s.t. 
}f(gh)=[\det{g}]^kf(h),\text{ }k\in{0,1,2,...}, \text{ }\forall g\in 
G,h\in H\}\nonumber.
\eqn
Clearly such invariants are relatively constant up to the determinant upon each orbit class of $H$. The set of invariants can, after some trivial definitions, be given the structure of a (graded) ring and it can be shown that there exists (under the action of the general linear group at least) a \textit{finite} set of elements which generate the full set on a given linear space. It can also be shown that the set of orbit classes of a given linear space can be completely classified given a full set of invariants on that space \cite{olver}. \\\indent
The motivation of the present work is the possibility that the study of orbit classes can be used to elucidate interesting results in phylogenetics.

\section{$K=2$ characters and qubits.}
\label{2characters}

From here on we specialize to the case where the set of characters consists of $K=2$ elements $\{0,1\}$. When we  are dealing with a single taxon the phylogenetic state $p$ mathematically corresponds to a vector belonging to $\mathbb{R}^2$. In quantum physics the corresponding two dimensional object is the "qubit" which in turn belongs to the vector space $\mathbb{C}^2$ and if we take multiple qubits the correct state space is $\mathcal{H}=\mathbb{C}^2\otimes\mathbb{C}^2\otimes ...\otimes\mathbb{C}^2$. As we showed previously the case of the phylogenetics of multiple taxa the corresponding state space is $H={\mathbb{R}^{2}\otimes\mathbb{R}^{2}\otimes ...\otimes\mathbb{R}^{2} }$. In the forgoing work we will be implicitly taking advantage of the fact that $\mathbb{R}\subset\mathbb{C}$.

\section{Canonical Forms}\label{canonical}

We wish to construct the orbit classes of $\mathcal{H}=\mathbb{C}^2\otimes\mathbb{C}^2$ under the group action $GL(\mathbb{C}^2)\times GL(\mathbb{C}^2)$. We have seen that for any state $h\in \mathcal{H}$ we can find an isomorphic state $\phi(h)\in L(\mathbb{C}^2)$ which transforms under the group action as $\phi\rightarrow\phi'=g_1\phi g_2^{\intercal}$. Hence we can answer the orbit class problem by taking a canonical $2\times2$ matrix $X$ and considering the set of matrices $M:M=AXB;A,B\in GL(\mathbb{C}^2)$.

\begin{theorem}
The vector space $V\otimes V$ where $V\equiv\mathbb{C}^2$ has three orbits under the group action $GL(V)\times GL(V)$. Under the isomorphism $V\otimes V\cong L(V)$ the orbits are characterized by the following canonical forms:
(i) Null-orbit $X=\left(\begin{array}{cc} 0 & 0 \\ 0 & 0\nonumber
\end{array}\right)
$;
(ii) Separable-orbit $Y=\left(\begin{array}{cc} 1 & 0 \\ 0 & 0\nonumber
\end{array}\right)
$;
(iii) Entangled-orbit $Z=\left(\begin{array}{cc} 1 & 0 \\ 0 & 1\nonumber
\end{array}\right)
$. The separable and entangled-orbits can be distinguished by the determinant function.
\end{theorem}

\begin{proof}
(i) The null-orbit has only one member, the null vector; which is of course unchanged by the group action.
(ii) We are required to show that the set of $2\times 2$ matrices $\mathcal{M}=\{S:S=AYB;A,B\in GL(V)\}$ is all matrices such that $\det(S)=0$. We begin by taking a general member of $\mathcal{M}$, $S=\left(\begin{array}{cc} a & b \\ c & d\nonumber
\end{array}\right)$ with $ad-bc=0$. Clearly the matrices 
\beqn
S':&=\left(\begin{array}{cc} 0 & 1 \\ 1 & 0
\end{array}\right)S=\left(\begin{array}{cc} c & d \\ a & b
\end{array}\right),\nonumber
\quad
S'':&=S\left(\begin{array}{cc} 0 & 1 \\ 1 & 0
\end{array}\right)=\left(\begin{array}{cc} b & a \\ d & c
\end{array}\right)\nonumber,\quad\mbox{and}
\eqn
\beqn
S''':&=\left(\begin{array}{cc} 0 & 1 \\ 1 & 0
\end{array}\right)S\left(\begin{array}{cc} 0 & 1 \\ 1 & 0
\end{array}\right)=\left(\begin{array}{cc} d & c \\ b & a
\end{array}\right)\nonumber
\eqn
also belong to $\mathcal{M}$. So without loss of generality we can take $a\neq 0$ and it is an easy computation to show that
\beqn
S=\left(\begin{array}{cc} 1 & 0 \\ c/a & 1\nonumber
\end{array}\right)Y\left(\begin{array}{cc} a & b \\ 0 & 1\nonumber
\end{array}\right),
\eqn
so that $\mathcal{M}$ is the set of $2\times 2$ matrices with vanishing determinant.
(iii) Clearly any $2\times 2$ matrix $N$ with non-zero determinant can be written as $N=AZB$ where $A,B\in GL(\mathbb{C}^2)$.\qed
\end{proof}
\begin{corollary}
The orbits of $\mathcal{H}=\mathbb{C}^2\otimes \mathbb{C}^2$ under $SL(\mathbb{C}^2)\times SL(\mathbb{C}^2)$ are labelled by the determinant function $\det[\phi(h)].$\end{corollary}
For further discussion see \cite{bern,lind,dur}.

\section{The concurrence}
\label{concurrencesection}

We consider the case of $L=2$ taxa derived from the branching of a single taxon at $t=0$ followed by arbitrary Markov evolution. The state is represented by a tensor in $H=\mathbb{R}^2\otimes \mathbb{R}^2$ and is expressed as
\beqn\label{L=2}
P(t)=M_{\epsilon_1}\otimes M_{\epsilon_2}\delta\cdot p,
\eqn 
where $t_1=t_2=t$, (see figure \ref{twotaxa}).\\\indent
The most general rate matrix depends on 2 parameters and can be expressed as
\beqn
R=\left(\begin{array}{cc} -\alpha & \beta \\ \alpha & -\beta\nonumber
\end{array}\right)
\eqn
where $\alpha$ and $\beta$ are real. A simple free parameter count in expression (\ref{L=2}) yields, taking into account the scaling symmetry on edges, 1 free parameter due to each transition matrix and 1 free parameter due to the initial state $p$. Hence there are a total of 3 free parameters and given that the components of the $K^2=4$ dimensional $P(t)$ are probabilites we conclude that all free parameters are accounted for.\\\indent
In quantum physics the tensors representing 2 qubits correspond in phylogenetics to the case of $L=2$ taxa with $K=2$ characters. As we have shown in the previous section, there exist 2 nontrivial orbit classes which are completely distinguished by the relative invariant known as the concurrence, $\mathcal{C}:\mathcal{H}=\mathbb{C}^2\otimes\mathbb{C}^2\rightarrow \mathbb{C}$. Using the formalism we have developed we can express the concurrence of the state $h\in \mathcal{H}$ as
\beqn\label{concurrence}
\mathcal{C}(h)=\det[\phi(h)],
\eqn
which satisfies
\beqn
\mathcal{C}(h'):&=\mathcal{C}(g_1\otimes g_2h)\nonumber\\
&=\det[g_1\phi(h)g_2^\intercal]\nonumber\\
&=\det[g_1]\det[g_2]\mathcal{C}(h),\nonumber
\eqn
so the concurrence is truly a relative invariant.
This can also be expressed explicitly as
\beqn\label{tensorconc}
\mathcal{C}(h)=\sum h_{ij}h_{kl}\epsilon_{ik}\epsilon_{jl},
\eqn
where $\epsilon$ is the completely anti-symmetric tensor with $\epsilon_{01}=1$.
The two orbit classes correspond to the completely entangled Bell state and the completely dis-entangled, and hence separable, state.  The entangled orbit is the set of states equivalent to the Bell state
\beqn
h_{bell}=\fra{1}{\sqrt{2}}(e_0\otimes e_0+e_1\otimes e_1),\nonumber
\eqn
whereas the dis-entangled orbit is the set of states which take on the separable form
\beqn
h=u\otimes v,\nonumber
\eqn
where $u,v\in \mathbb{C}^2$. The concurrence vanishes if and only if the state belongs to the separable orbit class. This property can be used to distinguish the orbit classes.\\\indent
In phylogenetic analysis the concurrence can be used to establish the magnitude of divergence between a pair of taxa derived from a single branching event. The case where there is no phylogenetic relation cannot be distinguished from the case of infinite divergence.  When there has been infinite divergence we have
\beqn
\lim_{t\rightarrow\infty}P(t)=\Pi=\pi_{\epsilon_1}\otimes\pi_{\epsilon_2},\nonumber
\eqn
which is a separable state and hence has concurrence 
\beqn
\mathcal{C}(\pi_{\epsilon_1}\otimes\pi_{\epsilon_2})=(\beta_1\beta_2)(\alpha_1\alpha_2)-(\beta_1\alpha_2)(\alpha_1\beta_2)=0.\nonumber
\eqn 
The concurrence of the phylogenetic state (\ref{L=2}) is given by
\beqn
\mathcal{C}(P(t))&=\det[M_{\epsilon_1}]\det[M_{\epsilon_2}]\det[\phi(\delta\cdot p)]\nonumber\\
\eqn
and using the operator identity $\det[e^X]=e^{trX}$ can easily be computed 
\beqn\label{concurrence1}
\mathcal{C}(P(t))&=\displaystyle{e^{tr[R_1t]}e^{tr[R_2t]}p_0p_1}\\
&=e^{-(\alpha_1+\beta_1+\alpha_2+\beta_2)t}p_0p_1.
\eqn
From this explicit form it can be seen that the concurrence is some kind of measure of phylogenetic divergence. From the concurrence we would like to construct a formal distance function. In the case of general $L$ taxa it is of course possible to construct a reduced tensor which represents the pattern distribution upon any pair of the taxa. This is achieved using the prescription defined in equation (\ref{reduced}). One can then go on to calculate the concurrence between any given pair of taxa taken from the set of $L$. We define a distance function, $d_{ij}$, between any pair of taxa taken from a set of $L$ as  
\beqn\label{distance}
d_{ij}:&=-\log{\mathcal{C}\left(\sum P_{a_1a_2...a_L}e_{a_i}\otimes e_{a_j}\right)},i\neq j\\
d_{ii}:&=0. 
\eqn
From the definition of the concurrence (\ref{tensorconc}) it is trivial to show that $d_{ij}=d_{ji}$. At the time, $\tau$, of the branching event at which the pair of taxa under consideration were created the concurrence took on the value $p_0p_1$ where 
\beqn\label{marginal}
p_\gamma=\sum_{\text{all $a$'s}} P_{a_1a_2...a_{i-1}\gamma a_{i+1}...a_{j-1}\gamma a_{j+1}...a_L}(\tau),
\eqn
Of course we have $0\leq p_0p_1\leq1$ and after this time the concurrence scales with the determinant $\det[M_i]\det[M_j]$ which is also strictly positive and less than unity. We can conclude that the concurrence between a pair of taxa is always strictly positive and less than unity and that the distance function, $d_{ij}$, is also strictly positive. The triangle inequality
\beqn
d_{ij}+d_{jk}\geq d_{ik}\nonumber
\eqn
 is equivalent to the statement that
\beqn
\mathcal{C}\left(\sum P_{a_1a_2...a_L}e_{a_i}\otimes e_{a_j}\right)\mathcal{C}&\left(\sum P_{a_1a_2...a_L}e_{a_j}\otimes e_{a_k}\right)&\nonumber\\&\leq \mathcal{C}\left(\sum P_{a_1a_2...a_L}e_{a_i}\otimes e_{a_k}\right),\nonumber
\eqn
which invoking (\ref{concurrence1}) can be expressed as
\beqn
e^{-(\alpha_i+\beta_i+\alpha_j+\beta_j)t}p^{(ij)}_0p^{(ij)}_1e^{-(\alpha_j+\beta_j+\alpha_k+\beta_k)t}&p^{(jk)}_0p^{(jk)}_1\nonumber\\&\leq e^{-(\alpha_i+\beta_i+\alpha_k+\beta_k)t}p^{(ik)}_0p^{(ik)}_1.\nonumber
\eqn
Here $p^{(ij)}$ is the single taxon marginal distribution existing at the node closest to the root which joins taxon $i$ to taxon $j$. These marginal distributions are calculated as in (\ref{marginal}).
Now depending on the branching structure of the tree we have $p^{(ij)}=p^{(jk)}$ or $p^{(jk)}=p^{(ik)}$ so that the distance function satisfies the triangle inequality.\\\indent
The distance function (\ref{distance}) is well known in phylogenetics as the $\log\det$ distance \cite{steel,fels}.

\section{The tangle}\label{tangle}

\begin{figure}[h]
\centering
\resizebox{0.3\hsize}{!}{\rotatebox{0}{\includegraphics{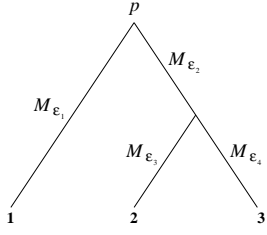}}}
\caption{The CTMC model of three taxa on the tree (1(23)).}
\label{threetaxa}
\end{figure}

We consider the case of $L=3$ taxa derived from the branching of a single taxon at $t=0$ followed by arbitrary Markov evolution, an additional branching event on edge 1 or 2 at $t=\tau$ and then additional arbitrary Markov evolution. For the case when the second branching event occurs on edge 2 the tree is represented by (1(23)) and the state is represented by a tensor in $H=\mathbb{R}^2\otimes \mathbb{R}^2\otimes \mathbb{R}^2$ as
\beqn\label{L=3}
P_\tau(t)=[M_{\epsilon_1}\otimes M_{\epsilon_3}\otimes M_{\epsilon_4}]1\otimes\delta [1\otimes M_{\epsilon_2}]\delta\cdot p,
\eqn
where $t_1\!=t$, $t_2=\tau$, $t_3=t_4=t-\tau$, (see figure [\ref{threetaxa}])\footnote{The use of pattern frequencies for the case of three taxa has been studied in relation to the problem of tree reconstruction by Pearl and Tarsi \cite{pearl} and Chang \cite{chang}.}.

It is known that there are 6 orbit classes of $\mathbb{C}^2\otimes \mathbb{C}^2\otimes \mathbb{C}^2$ under $GL(\mathbb{C}^2)\times GL(\mathbb{C}^2)\times GL(\mathbb{C}^2)$ that can be distinguished by functions of the concurrence and another relative invariant known as the tangle \cite{dur,guhn}. We begin by defining a partial concurrence operation $\{\mathcal{C}_a:\mathbb{C}^2\otimes \mathbb{C}^2\otimes \mathbb{C}^2\rightarrow \mathbb{C}^2\otimes \mathbb{C}^2,a=1,2,3.\}$ as
\beqn\label{pconcurrence}
\mathcal{C}_1(h)=\sum h_{ijk}h_{lmn}\epsilon_{jm}\epsilon_{kn}e_i\otimes e_l,\\
\mathcal{C}_2(h)=\sum h_{ijk}h_{lmn}\epsilon_{il}\epsilon_{kn}e_j\otimes e_m,\\
\mathcal{C}_3(h)=\sum h_{ijk}h_{lmn}\epsilon_{il}\epsilon_{jm}e_k\otimes e_n.\\
\eqn
From these definitions it is easy to see that
\beqn
\mathcal{C}_1(h'):&=\mathcal{C}_1(g_1\otimes g_2 \otimes g_3 h)\nonumber\\
&=[\det(g_2)\det(g_3)]g_1\otimes g_1\mathcal{C}_1(h),
\eqn
with similar expressions for $\mathcal{C}_2$ and $\mathcal{C}_3$. The tangle, $\{\mathcal{T}:\mathbb{C}^2\otimes \mathbb{C}^2\otimes \mathbb{C}^2\rightarrow \mathbb{C}\}$, can be defined as
\beqn
\mathcal{T}=\mathcal{C}\raisebox{.3ex}{\scriptsize o} \mathcal{C}_a,\nonumber
\eqn
where we will confirm shortly that $\mathcal{T}$ is independent of the choice of $a$. The tangle is a relative invariant satisfying
\beqn
\mathcal{T}(h')=[\det(g_1)\det(g_2)\det(g_3)]^2\mathcal{T}(h),
\eqn
and, in analogy to \ref{tensorconc},  can also be written in the form
\beqn
\mathcal{T}(h)=\sum h_{a_1a_2a_3}h_{b_1b_2b_3}h_{c_1c_2c_3}h_{d_1d_2d_3}\epsilon_{a_1b_1}\epsilon_{b_2c_1}\epsilon_{c_2d_1}\epsilon_{d_2a_2}\epsilon_{b_2d_3}\epsilon_{a_3c_3}.
\eqn
\\\indent
The 6 orbit classes are described by the completely dis-entangled states 
\beqn
h=&u\otimes v\otimes w,\\\nonumber u&,v,w\in \mathbb{C}^2;
\eqn
the partially entangled states
\beqn
h_a:\quad a=1,2,3,\nonumber
\eqn
 which form 3 orbit classes characterized by the separability of the canonical tensors
\beqn
\text{(1) }h_{ijk}=u_iv_{jk},\\\nonumber
\text{(2) }h_{ijk}=u_{ij}v_k,\\\nonumber
\text{(3) }h_{ijk}=u_{ik}v_j;\nonumber
\eqn
the completely entangled states equivalent to the $GHZ$ state
\beqn
h_{ghz}=\fra{1}{\sqrt{2}}(e_0\otimes e_0\otimes e_0+e_1\otimes e_1\otimes e_1);\nonumber
\eqn
and the completely entangled states equivalent to the $W$ state
\beqn
h_w=\fra{1}{\sqrt{3}}(e_0\otimes e_0\otimes e_1+e_0\otimes e_1\otimes e_0+e_1\otimes e_0\otimes e_0).\nonumber
\eqn
The tangle and the concurrence and its partial counterparts can be used to fully distinguish these orbit classes.  For the completely dis-entangled tensors we have
\beqn
\mathcal{C}_a(h)=0,\quad\forall a;\nonumber
\eqn 
whereas for the partially entangled states we have
\beqn
\mathcal{C}_a(h_{a'})=0,\quad\text{ iff }\delta_a^{a'}=0.\nonumber
\eqn
The $GHZ$ and $W$ orbits are distinguished by calculating the tangle
\beqn
\mathcal{T}(h_{ghz})&\neq 0,\quad\mathcal{T}(h_W)&=0.\nonumber
\eqn
From these properties we will now show that the tangle is indeed independent of the choice of the partial concurrence. Begin by introducing an action of the symmetric group, $S_3$, on the tensor product space defined by 
\beqn
\sigma\in S_3:V_1\otimes V_2\otimes V_3\rightarrow V_{\sigma 1}\otimes V_{\sigma 2}\otimes V_{\sigma 3}.\nonumber
\eqn
Now since the tangle vanishes everywhere except on the $GHZ$ orbit and $\sigma h_{ghz}=h_{ghz}$ we need only consider the value of the tangle on elements which lie on the $GHZ$ orbit. We take as our element $x=g_1\otimes g_2\otimes g_3h_{ghz}$ and proceed. From its definition the tangle satisfies
\beqn
\mathcal{T}(\sigma h)=\mathcal{C}\raisebox{.3ex}{\scriptsize o} \mathcal{C}_a(\sigma h)&=\mathcal{C}\raisebox{.3ex}{\scriptsize o} \mathcal{C}_{\sigma a}(h),\nonumber\\
\eqn
$\forall h\in \mathcal{H}$. For our element $x$ we have
\beqn
\mathcal{T}(\sigma x)&=\mathcal{T}(g_{\sigma 1}\otimes g_{\sigma 2}\otimes g_{\sigma 3}h_{ghz}),\nonumber\\
&=[\det{g}]^2\mathcal{T}(h_{ghz}),\nonumber\\
&=\mathcal{T}(x).
\eqn
This shows that
\beqn
\mathcal{C}\raisebox{.3ex}{\scriptsize o} \mathcal{C}_a&=\mathcal{C}\raisebox{.3ex}{\scriptsize o} \mathcal{C}_{\sigma a},\nonumber\\
\forall \sigma&\in S_3.
\eqn
\\\indent
We now determine which orbit the $L=3$ phylogenetic state (\ref{L=3}) lies in. The easiest way to do this is to calculate the various invariants at the time of the branching event. To this end we use (\ref{comp}) so that components of the state are
\beqn\label{bulldust}
p_{i_1i_2i_3}(\tau)=p_{i_1i_2}(\tau)\delta_{i_2i_3}.
\eqn
From this expression we might expect that the entanglement of the state given by the tangle just after branching can be expressed as a function of the entanglement in the state given by the concurrence just before branching. This is indeed the case. By direct computation using (\ref{bulldust}) it can be shown that at the time of branching the tangle is given by 
\beqn
\mathcal{T}(P_{\tau}(\tau))&=\mathcal{C}\raisebox{.3ex}{\scriptsize o}\mathcal{C}_3(P_{\tau}(\tau))\nonumber\\
&=-2[\mathcal{C}(M_{\epsilon_1'}\otimes M_{\epsilon_2}\delta\cdot p)]^2,\nonumber\\
\eqn
where $\epsilon_1'=\{\alpha_1,\beta_1;t_2\}$.
The tangle has the value
\beqn
\mathcal{T}(P_{\tau}(\tau))=-2e^{-2(\alpha_1+\beta_1+\alpha_2+\beta_2)\tau}[p_0p_1]^2.
\eqn
 Subsequent to this the tangle takes on the value
\beqn
\mathcal{T}(P_{\tau}(t))&=(\det[M_{\epsilon_1''}]\det[M_{\epsilon_3}]\det[M_{\epsilon_4}])^2\mathcal{T}(P_{\tau}(\tau))\nonumber\\
&=-2e^{-2(\alpha_1+\beta_1+\alpha_3+\beta_3+\alpha_4+\beta_4)(t-\tau)}e^{-2(\alpha_1+\beta_1+\alpha_2+\beta_2)\tau}(p_0p_1)^2\nonumber,
\eqn
so that the phylogenetic state belongs to the $GHZ$ orbit for all finite $t$. This equivalence has in fact been observed in a different context in \cite{lake}.
It should be noted that as $t\rightarrow\infty$ the tangle tends to zero and the state becomes the disentangled, stationary state which corresponds phylogenetically to the case where the taxa are unrelated. This is of course what we would expect if the taxa have diverged so much that there is no longer a possibility of establishing any relation between the taxa.\\\indent

\section{The tangle and distance functions}\label{distances}

The tangle gives us a new tool for calculating the phylogenetic distance between a set of three taxa. As was the case with the concurrence it is possible to calculate the value of the tangle for any subset of three taxa taken from a set of $L$ taxa. We use the tangle to define a three taxa phylogenetic distance  given by
\beqn
d_{ijk}:=\fra{1}{2}\log{2}-\fra{1}{2}\log{\left[-\mathcal{T}\left(\sum P_{a_1a_2...a_L}e_{a_i}\otimes e_{a_j}\otimes e_{a_k}\right)\right]}.
\eqn
For the case under consideration this three taxa distance takes on the value
\beqn
d_{123}=(\alpha_1+\beta_1)&t+(\alpha_2+\beta_2)\tau\nonumber\\&+(\alpha_3+\beta_3+\alpha_4+\beta_4)(t-\tau)-\log{p_0p_1}.\nonumber
\eqn
\\\indent
At this point it is illuminating to compare the values of all the partial distance functions $d_{12},d_{23}$ and $d_{13}$ with the value of the tangle for the case of the branching structure $(1(23))$. The distance functions take on the values 
\beqn
d_{12}&=(\alpha_1+\beta_1)t+(\alpha_2+\beta_2)\tau +(\alpha_3+\beta_3)(t-\tau)-\log{p_0p_1},\nonumber\\
d_{23}&=(\alpha_3+\beta_3+\alpha_4+\beta_4)(t-\tau)-\log{p_0'p_1'},\nonumber\\
d_{13}&=(\alpha_1+\beta_1)t+(\alpha_2+\beta_2)\tau +(\alpha_4+\beta_4)(t-\tau)-\log{p_0p_1},
\eqn 
where $p'=M_{\epsilon_2}p$.
We define weights on the edges $1,2,3,4$ of the tree to be 
\beqn
x&=(\alpha_1+\beta_1)t,\nonumber\\
y&=(\alpha_2+\beta_2)\tau,\nonumber\\
z&=(\alpha_3+\beta_3)(t-\tau),\nonumber\\
w&=(\alpha_4+\beta_4)(t-\tau),\nonumber\\
\eqn
respectively.
It is possible to solve the distance function equations for the weights
\beqn
x+y&=d_{12}+d_{13}-d_{123}+\log{p_0p_1},\nonumber\\
z&=d_{123}-d_{13},\nonumber\\
w&=d_{123}-d_{12},\nonumber\\
\log{p_0'p_1'}&=2d_{123}-d_{12}-d_{13}-d_{23}.
\eqn
In summary we find that, if we assume the branching structure of the tree, we now have a prescription that gives us the evolutionary distances between three taxa up to errors caused by the fact we cannot determine the marginal distribution, $p$, which lies the at top node of the tree. This marginal distribution must be estimated using some resonable prescription.  \\\indent
To elucidate the value of including the tangle in the analysis we present the corresponding set of branch lengths calculated without using the tangle
\beqn
x+y&=\fra{1}{2}(d_{12}+d_{13}-d_{23})-\log{p_0p_1}-\fra{1}{2}\log{p'_0p'_1},\nonumber\\
z&=\fra{1}{2}(d_{12}+d_{23}-d_{13})+\fra{1}{2}\log{p_0'p_1'},\nonumber\\
w&=\fra{1}{2}(d_{23}+d_{13}-d_{12})+\fra{1}{2}\log{p_0'p_1'}.
\eqn
\\\indent
This comparison makes clear the advantage of including the tangle in the analysis of branch lengths.

\section{Conclusion}

We have shown that it is possible to present the continuous time Markov chain model of phylogenetics using a dynamical, tensor state space description. We have shown that the branching process introduces entanglement into the description, and that the group invariant approach to entanglement in quantum physics can be used in phylogenetics to derive distance functions between taxa. The main original result presented was the use of the tangle as a new distance measure between three taxa.\\\indent
Entanglement measures can be extended to the cases of $K>2$ and $L>3$ and will be explored in future work. In particular the invariant theory for the case of $K=2,L=4$ is established in the physics literature \cite{vers,luqu} and progress in interpreting the theory in the phylogenetic context is underway.\\\indent The use of invariant functions to distinguish alternate tree branching structures was not achieved in this work, but future work will explore sharpening the group action from $GL(V)^{\times L}$ to the more stringent Markov operator action and it is hoped that this will allow branching structures to be distinguished using the corresponding invariant functions. This will allow results establishing that if and only if the values of invariant functions taken on an arbitrary character distribution satisfy certain relations, then it can be concluded that the distribution was generated from a Markov model on a given branching structure.\\\indent
There is the remaining issue that the distance functions are known only up to $\log$ functions of the marginal distributions at the nodes. An avenue of furthur research would be to determine under what conditions these $\log$ functions can be treated as statistically insignificant. This would involve studying how far the sequences can diverge before the information content of the distribution becomes so small that the calculation of edge weights is misleading. That is, under what circumstances are the edge weights large enough so that terms such as $\log{p_0p_1}$ can be treated as small error? The problem is that the edge weights are a measure of the magnitude of divergence, which if allowed to become large enough means that the information content of the distribution is small, and hence there is no potential to establish phylogenetic relation anyway. It would be fruitful to explore these issues analytically.\\\indent

\begin{acknowledgement}
PDJ and JGS thank the Department of Physics and Astronomy, and also
the Biomathematics Research Centre, University of Canterbury, Christchurch, New 
Zealand, for hosting a visit during which this work was presented, clarified and expanded upon.
We would also like to thank Mike Steel and Jim Bashford for helpful comments. Finally, we would like to the thank the anomynous referrees for helpful comments regarding the initial draft which have led to a much improved text in the final version and, also, John Rhodes for pointing out an error in the edge weights analysis. This research was supported by the Australian Research Council  
grant DP0344996 and the Australian Postgraduate Award.
\end{acknowledgement}

\end{document}